\documentclass{article}
\usepackage{amsmath,graphicx,mlspconf}

\usepackage{booktabs}
\usepackage{soul} 
\usepackage{amsmath,amsfonts,amssymb,amsbsy}
\usepackage{url}
\usepackage{xcolor}
\usepackage{footnote}
\usepackage{graphicx} %
\usepackage{caption}
\usepackage{subcaption}
\usepackage{placeins} %
\usepackage[titletoc]{appendix} %
\usepackage[colorlinks,linkcolor=black,citecolor=black,urlcolor=blue,filecolor=blue,backref=page]{hyperref}

\usepackage{ifthen}
\usepackage{tikz,pgfplots}
\usetikzlibrary{matrix}
\usetikzlibrary{calc}
\newlength{\figurewidth}
\newlength{\figureheight}

\def\figpdfdir{./} %
\def\figtikzdir{./} %

\newcommand{\minput}[2][]{
\ifthenelse{\equal{#1}{pdf}}
	{ \includegraphics{\figpdfdir #2} }
	{ \tikzset{external/remake next} \tikzsetnextfilename{#2} \input{\figtikzdir #2} }
}

\usetikzlibrary{external}
\tikzset{external/system call={lualatex
	\tikzexternalcheckshellescape -halt-on-error -interaction=batchmode
	-jobname "\image" "\texsource"}}
\newcommand{\vc}[1] { \mathbf{#1} }
\newcommand{\vs}[1] { \boldsymbol{#1} }
\newcommand{\tp}{\mathsf{T}}
\newcommand{\ti}[1] { \tilde{#1} } 
 
\newcommand{\tx}[1] { \text{#1} } 
\newcommand{\given} { \,|\, }

\newcommand{\tr}[1] { \mathrm{Tr} {\left(#1\right)} }
\newcommand{\KL}[2] { \mathrm{KL} {\left(#1 \, \| \, #2\right)} }
\newcommand{\bKL}[2] { \mathrm{KL} {\left(#1 \, \big\| \, #2\right)} }

\newcommand{\Unif}[1]{ \mathrm{U} {\left(#1\right)} }
\newcommand{\Normal}[1] { \mathrm{N} {\left(#1\right)}  }

\newcommand{\subjto}{ \, \tx{s.t.} }
\title{PROJECTION PREDICTIVE MODEL SELECTION FOR GAUSSIAN PROCESSES}
\name{Juho Piironen, Aki Vehtari \thanks{We acknowledge the computational resources provided by the Aalto Science-IT project.}}
\address{Helsinki Institute for Information Technology HIIT, \\ Department of Computer Science, Aalto University}
\begin{document}

\maketitle
\begin{abstract}
We propose a new method for simplification of Gaussian process (GP) models by projecting the information contained in the full encompassing model and selecting a reduced number of variables based on their predictive relevance.  Our results on synthetic and real world datasets show that the proposed method improves the assessment of variable relevance compared to the automatic relevance determination (ARD) via the length-scale parameters.  We expect the method to be useful for improving explainability of the models, reducing the future measurement costs and reducing the computation time for making new predictions.
\end{abstract}
\begin{keywords}
Gaussian processes, ARD, model selection, variable selection
\end{keywords}

\section{Introduction}
\label{sec:intro}

Gaussian processes (GPs) offer a flexible, yet tractable way of defining priors over latent functions \cite{rasmussen2006book}.
We are often not only interested in learning the mapping from the predictor variables to the target variable, but also finding out, which of the variables are relevant for the model. 
The relevance of a variable is naturally defined by its predictive power on the target variable \cite{vehtari2012}.
Assessment of variable relevances is beneficial for improving the model
interpretability and for reducing the required measurements in the future if
the measurements are costly or otherwise difficult to obtain.  The
applications include, among others, medical studies, such as biomarker
selection for disease risk prediction \cite{peltola2014}, and various
industrial applications.

In this paper we consider the projection predictive variable selection
which was originally proposed for parametric generalized linear models
\cite{goutis1998,dupuis2003}. The novelty of this paper is extending
the projection predictive method for non-parametric Gaussian processes.
The projection predictive method works by first building a reference
model using all the available covariates and possibly a sparsifying
prior such as spike-and-slab \cite{mitchell1988,george1993},
hierarchical shrinkage \cite{piironen2016} or automatic relevance determination
\cite{MacKay:1994a,Neal:1996a}, and then solving
the decision problem how to make optimal predictions in the future if
only part of the variables are then available. Using all available
training data variables in the reference model and then projecting the
posterior information to the smaller restricted model, produces better
models than creating models only with the variable subset available in
the future \cite{piironen2016}. That is, we are able to exploit the
information from the all variables available during the model
construction phase even if they are not all available later when
making new predictions.

We also discuss and demonstrate why the so called automatic relevance
determination (ARD) hierarchical prior for separate length-scale
parameters \cite{MacKay:1994a,Neal:1996a} is
misleading in assessing variable relevancies. ARD parameters are related
to the non-linearity and thus it is difficult to differentiate whether
a short length-scale indicates strong predictive relevance or merely a
non-linearity of the latent function with respect to that dimension
(with possibly a small magnitude).
Our experiments show that the proposed method gives an improved and more robust performance compared to ARD, being able to better retain the predictive performance of the full model with a restricted number of variables.

In the Bayesian literature for linear models, a widely used idea is to
employ the spike-and-slab prior \cite{mitchell1988,george1993} which
gives a nonzero prior probability for an variable effect being exactly
zero. The relevance of the variables is then assessed based on their
marginal posterior probabilities to be non-zero. Although
spike-and-slab priors have been used also for GPs
\cite{vehtari2001thesis,savitsky2011}, we leave them
out of this study due to the space constraints. Furthermore, Piironen
and Vehtari \cite{piironen2016} demonstrated that marginal posterior
probabilities based variable selection is not as effective as the
projection predictive approach.

We first review the idea of using ARD prior for assessing variable
relevancies and discuss and demonstrate the problem with that.  We
then present our new method and discuss its connections to existing
literature.  Finally, we show some results using synthetic and real
world data which demonstrate the benefits of the proposed method.

\section{Background}
\label{sec:background}

In this section we will briefly go through our notation for Gaussian processes and discuss the idea behind the automatic relevance determination for assessing the input relevancies, and motivate the possible problem.

\subsection{Gaussian processes}

Gaussian processes (GP) \cite{rasmussen2006book} are a flexible nonparametric model in which the prior is specified directly on the latent function $f(\vc x)$, where $\vc x$ is the $D$-dimensional input vector.
The prior assumptions are encoded in the covariance function $k(\vc x, \vc x')$, which defines the covariance between two latent function values $f(\vc x)$ and $f(\vc x')$.
The standard zero mean GP prior is given by
\begin{align*}
	p(\vc f) = \Normal{\vc f \given 0, \vc K},
\end{align*}
where $\vc K$ denotes the covariance matrix between the latent function values $\vc f$ at the training inputs $\vc X=(\vc x_1^\tp,\dots,\vc x_n^\tp)$, so that $\vc K_{ij} = k(\vc x_i, \vc x_j)$.
Assuming Gaussian observation model for the target values $\vc y$ given the latent values and the noise variance $\sigma^2$, we can write the joint distribution of the observed targets $\vc y$ and the latent values $\vc f_*$ at test inputs $\vc X_*$ as
\begin{align*}
	\begin{bmatrix}	\vc y \\ \vc f_* \end{bmatrix}
	\sim \Normal{0, \begin{bmatrix} \vc K + \sigma^2\vc I & \vc K_*^\tp \\ 
									\vc K_* & \vc K_{**} \end{bmatrix}}.
\end{align*}
Here $\vc K_*$ denotes the covariance between the latent values at the training and test inputs, and $\vc K_{**}$ the covariance at the test inputs.
Using the Gaussian conditioning rule, we get the predictive distribution for $\vc f_*$ given $\vc y$
\begin{align}
\begin{split}
	\vc f_* \given \vc y &\sim \Normal{\vc f_* \given \vs \mu_*, \vs \Sigma_*}, \\
	\vs \mu_* &= \vc K_*(\vc K + \sigma^2 \vc I)^{-1} \vc y \\
	\vs \Sigma_* &= \vc K_{**} - \vc K_{*}(\vc K + \sigma^2 \vc I)^{-1}\vc K_{*}^\tp.
\label{eq:gp_pred}
\end{split}
\end{align}

\subsection{Automatic relevance determination}
\label{sec:ard}

The selection of the covariance function $k(\vc x, \vc x')$ has a great influence on the inference results and predictions of the GP model.
A popular choice and the one used this paper is the squared-exponential (SE) or exponentiated quadratic covariance function
\begin{align}
	k_\tx{SE}(\vc x,\vc x') 
	= \sigma_f^2\exp\left(-\frac{1}{2}\sum_{j=1}^D \frac{(x_j-x_j')^2}{\ell_j^2} \right),
	\label{eq:se_kernel}
\end{align}
which produces smooth functions.
Here each of the $D$ input dimensions is given a different length-scale parameter $\ell_j$ which allow the function to vary at different speed with respect to different inputs, while $\sigma_f^2$ defines the overall magnitude for the variability.

The use of separate length-scales is often referred to as the automatic relevance determination (ARD).
The intuition is that the marginal likelihood would favor solutions with large length-scales for those inputs along which the latent function is essentially flat.
The ARD value is then defined as the inverse of the length-scale which gives a ``relevance'' measure for each input.
However, while an infinitely long length-scale implies that the latent function is constant along that input dimension, finite ARD values are much more problematic to interpret.
The problem is that the inputs with nonlinear effect will have shorter length-scales than those with linear effect.

\begin{figure}[t]
\centering
	\setlength{\figureheight}{0.2\textwidth}
	\setlength{\figurewidth}{0.45\textwidth}
	\pgfplotsset{
	compat=newest,
	major tick length={0.05cm},
	x tick label style={font=\tiny},
	y tick label style={font=\tiny},
	} 
	\minput[pdf]{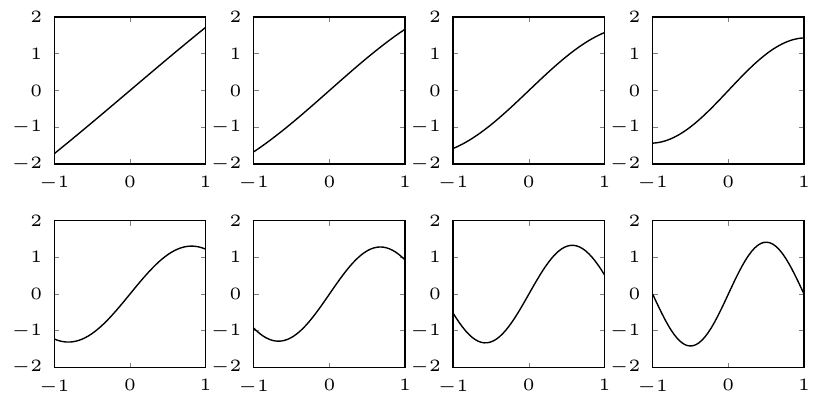}
	\caption{Latent functions $f_j(x_j)$, $i=1,\dots,8$ for the ARD experiment. Each function has a unit variance over the distribution of the corresponding input.}
	\label{fig:sines}
\end{figure}

This is easily illustrated with the following toy example:
\begin{align*}
	y &= f_1(x_1) + \dots + f_8(x_8) + e,\\
	\quad f_j(x_j) &= A_j \sin(\phi_j x_j), \quad j=1,\dots,8, \\	
	\quad x_j &\sim \Unif{-1,1}, \\ \quad e &\sim \Normal{0,{0.3}^2}.
\end{align*}
The coefficients $\phi_j$ are taken from an evenly spaced grid from $\pi/10$ to $\pi$ and the amplitudes $A_j$ are chosen so that the variance of each $f_j$ is one (over the corresponding $x_j$) making the inputs equally relevant in the squared error sense.
The functions $f_j$ are plotted in Figure~\ref{fig:sines}.

\begin{figure}
	\centering 
	\setlength{\figureheight}{0.15\textwidth}
	\setlength{\figurewidth}{0.45\textwidth}
	\pgfplotsset{
	compat=newest,
	major tick length={0.05cm},
	legend style={font=\footnotesize}, 
	x tick label style={font=\footnotesize},
	y tick label style={font=\footnotesize},
	} 
	\minput[pdf]{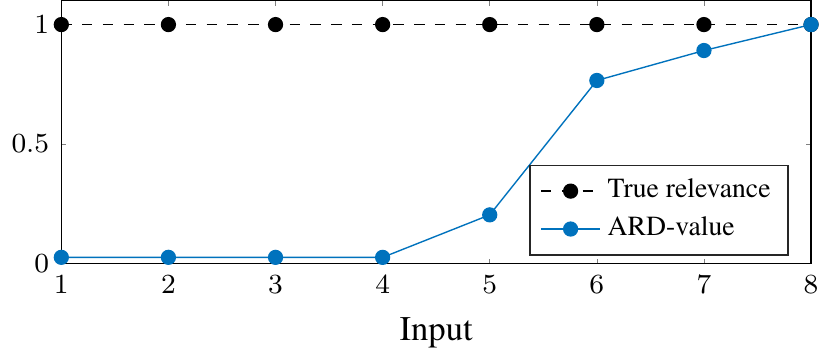} 
	\caption{ARD values (inverses of the length-scales) for the different inputs of the trained GP, averaged over 200 data realizations. The values are scaled so that the highest value is one.}
	\label{fig:ard1}
\end{figure}

We constructed a GP model with a constant and SE covariance functions \eqref{eq:se_kernel}  and $n=300$ training points.
The hyperparameter values were chosen by maximizing the marginal likelihood.
Figure~\ref{fig:ard1} shows the optimized ARD values (inverses of the length-scales), averaged over 200 different data realizations.
In terms of the ARD values, the inputs with almost linear response are estimated to be far less important than the nonlinear ones, although all of them are equally relevant.
This example shows that a small length-scale is related to the nonlinearity of the response rather than high predictive relevance.

We note that our intention is not to criticize using separate length-scales for different inputs.
Allowing individual lengthscales for the inputs increases the flexibility of the model and often improves the fit compared to using the same length-scale for all the inputs.
We are merely pointing out that, in general, the length-scale is an unreliable indicator of predictive relevance due to the tendency of favoring nonlinear inputs.
In Section~\ref{sec:toy_example_revisited} we demonstrate that our new method is not liable to the this kind of behaviour.

\section{Projection framework for \\Gaussian processes}

This section describes our proposed method for constructing and comparing GP submodels by utilizing the information in the reference GP model.
We first review the original approach for parametric linear models and then extend this to nonlinear GPs.

\subsection{Projection for generalized linear models}
\label{sec:projection_glm}

\subsubsection{General method}
\label{sec:projection_framework}

The idea in the predictive projection \cite{goutis1998,dupuis2003} is to simplify the full model $M$ (e.g. encompassing model with all covariates) by projecting the information in the posterior onto the simpler submodels so that the predictions change as little as possible. The change in the predictive distribution is measured by Kullback-Leibler divergence.
Recently Piironen and Vehtari \cite{piironen2016} demonstrated the superiority of the approach compared to many other Bayesian model selection methods for variable selection in generalized linear models.

Given the parameters $\vs \theta$ of the full model $M$, the projected parameters $\vs \theta^\perp$ of the submodel $M_\perp$ are solved from
\begin{align}
	\min_{\vs \theta^\perp}. \quad
	\frac{1}{n}\sum_{i=1}^n \KL{p(y_i \given \vs \theta, M)} {p(y_i \given \vs \theta^\perp, M_\perp)} \,.
\label{eq:projection}
\end{align}
The minimum of the above function (w.r.t. $\vs \theta^\perp$) is then defined to be the projection error $\delta(\vs \theta,M_\perp)$ for this particular parameter value $\vs \theta$.
Given a sample $\{\vs \theta_s\}_{s=1}^S$ from the posterior of the full model, one can compute the projected parameters $\{\vs \theta_s^\perp\}_{s=1}^S$ individually according to~\eqref{eq:projection}, and compute the overall projection error as the average divergence over the samples $\delta(M_\perp) = \frac{1}{n}\sum_{s=1}^S \delta(\vs \theta_s,M_\perp)$.
In model selection, one then seeks for a submodel $M_\perp$ with minimal projection error given the submodel complexity, for instance, the number of inputs we can afford to use.

\subsubsection{Linear Gaussian model}
\label{sec:projection_lgm}

Consider the model
\begin{align}
\begin{split}
	f_i &= \vc w^\tp \vc x_i \\
	y_i &= f_i + \varepsilon_i, \quad \varepsilon_i \sim \Normal{0,\sigma^2}, \quad i=1,\dots,n %
\end{split}
\label{eq:lgm}
\end{align}
where $\vc x$ is the $D$-dimensional vector of inputs, $\vc w$ contains the corresponding weights and $\sigma^2$ is the noise variance.
Given a sample $\vs \theta=(\vc w,\sigma^2)$ from the posterior of the full model, the projected parameters $\vs \theta_\perp=(\vc w_\perp,\sigma_\perp^2)$ for a submodel with fewer inputs can be solved analytically from Eq.~\eqref{eq:projection}.
Writing down the analytic form of the KL-divergence between two Gaussians, it is straightforward to show that the projected weights $\vc w_\perp$ are solved from
\begin{align}
\begin{split}
	\min_{\vc w_\perp}. \quad &(\vc f- \vc f_\perp)^\tp (\vc f- \vc f_\perp), \\ 
	\subjto \quad & \vc f = \vc X \vc w, \\
	 				& \vc f_\perp = \vc X_\perp \vc w_\perp.
\end{split}
\label{eq:projection_minimization}
\end{align}
Here $\vc X = (\vc x_1^\tp,\dots,\vc x_n^\tp)$ denotes the $n\times D$ predictor matrix of the full model, and $\vc X_\perp$ the contains those columns of $\vc X$ that correspond to the submodel we are projecting onto.
The solution is given by
\begin{align}
	\vc w_\perp = ({\vc X_\perp}^\tp \vc X_\perp)^{-1} {\vc X_\perp}^\tp  \vc f.
\label{eq:proj_weight}
\end{align}
Given the projected weights, the optimal noise variance of the projected models is then
\begin{align}
	\sigma^2_\perp = \sigma^2 + \frac{1}{n}(\vc f- \vc f_\perp)^\tp (\vc f- \vc f_\perp).
\label{eq:proj_noise}
\end{align}

The projection equations \eqref{eq:proj_weight} and \eqref{eq:proj_noise} have a natural interpretation.
The projected weights are determined by the least squares solution with the observations $\vc y$ replaced by the fit of the full model $\vc f = \vc X \vc w$.
The projected noise variance is the noise level of the full model plus the mismatch between the latent values of the full and the projected model.

\subsection{Projection for Gaussian processes}

\subsubsection{Gaussian observation model}
\label{sec:gpproj_gaussian}

Provided that we have fitted the full GP model, we have the predictive mean and covariance for the latent function, Eq.~\eqref{eq:gp_pred}.
The projection method designed for parametric models (Sec.~\ref{sec:projection_glm}) cannot be directly applied, because the parameters of the GP are essentially the latent values $\vc f$ and without constraints for the latent values in the submodel, the solution to the minimization problem \eqref{eq:projection_minimization} is $\vc f_\perp = \vc f$.

We propose a new method to perform the projection for GPs by requiring that the submodel prediction satisfies the predictive equations~\eqref{eq:gp_pred} and then minimize the posterior latent KL-divergence at training points with respect to the hyperparameters of the submodel
\begin{align}
\begin{split}
	\min_{\vs \theta_\perp}. \quad & \bKL{\Normal{\vc f\given \vs \mu, \vs \Sigma}} 
								{ \Normal{\vc f\given \vs \mu_\perp, \vs \Sigma_\perp} }, \\ 
	\subjto \quad & \vs \mu_\perp = \vc K_\perp(\vc K_\perp + \sigma^2 \vc I)^{-1} \vc y, \\
	 				& \vs \Sigma_\perp = \vc K_\perp - \vc K_\perp(\vc K_\perp + \sigma^2 \vc I)^{-1}\vc K_\perp,
\end{split}
\label{eq:gpproj_problem}
\end{align}
where $\vs \mu$ and $\vs \Sigma$ are the mean and covariance for the latent values at the training inputs in the full model, and $\vc K_\perp$ the training covariance matrix of the submodel with hyperparameters $\vs \theta_\perp$.
Note here that $\sigma^2$ denotes the noise variance of the {\it full} model; the same noise variance is used for the submodel.
To account for the increased uncertainty regarding the latent function when moving from the full model to the submodel (e.g. when removing inputs), we include an additional diagonal noise variance component $\sigma^2_0 \vc I$ in the submodel covariance $\vc K_\perp$.
This corresponds to the increased noise variance in projection for the linear Gaussian model, Eq.~\eqref{eq:proj_noise}, where the increased uncertainty is added to the latent function.
At extreme, when the submodel is the null model (no variables), then $\vc K_\perp = \sigma^2_0 \vc I$ and the whole signal is explained by noise.

When comparing submodels, we define the projection error to be
\begin{align}
	\delta(M_\perp) = \bKL{\Normal{\vc f\given \vs \mu, \vs \Sigma}} 
					{ \Normal{\vc f\given \vs \mu_\perp', \vs \Sigma_\perp'} },
\label{eq:pred_divergence}
\end{align}
where $\vs \mu'_\perp$ and $\vs \Sigma'_\perp$ are the {\it predictive} mean and covariance for the latent function at the training locations
\begin{align}
	\vs \mu'_\perp &= \vc K_\perp'(\vc K_\perp + \sigma^2 \vc I)^{-1} \vc y, \\
	\vs \Sigma'_\perp &= \vc K_\perp - \vc K_\perp'(\vc K_\perp + \sigma^2 \vc I)^{-1}\vc K_\perp'.
\end{align}
The difference between these and $\vs \mu_\perp$ and $\vs \Sigma_\perp$ in Eq.~\eqref{eq:gpproj_problem} is that here the covariance between the training and ``test'' points,  $\vc K_\perp'$, does not contain the additional noise variance $\sigma^2_0 \vc I$.
In other words, $\vc K_\perp' = \vc K_\perp - \sigma^2_0 \vc I$.

One can see the difference most easily by considering the null model $\vc K_\perp = \sigma^2_0 \vc I$.
In this case, the variation of the latent function is modeled by independent noise, and the covariance between the latent values at the observed and future points is zero $\vc K'_\perp=0$ even if consider the prediction at the training points.
Assuming that the noise variance $\sigma^2$ is fairly small compared to the variance of the latent function $\sigma^2_0$, there is a clear difference between the predictive mean $\vs \mu_\perp' = 0$ (zero everywhere) and the posterior mean $\vs \mu_\perp \approx \vc y$ (close to the observed targets).
For this reason, the predictive divergence \eqref{eq:pred_divergence} appears as a more natural measure for projection error than the posterior divergence used to optimize the hyperparameters \eqref{eq:gpproj_problem}.

A natural question then follows, why not to use the predictive divergence \eqref{eq:pred_divergence} as the objective function for learning the submodel hyperparameters?
The problem with this approach is that when $\vs \mu$ is close to $\vc y$ (i.e., noise variance $\sigma^2$ is fairly small) the divergence~\eqref{eq:pred_divergence} can be made small by interpolating through the observed targets $\vs \mu'_\perp = \vc y$, which is obtained by letting the length-scales of the submodel to go very small and the magnitude very large (both $\vc K_\perp$ and $\vc K_\perp'$ become almost diagonal, with very large values, so that $\vc K_\perp'(\vc K_\perp + \sigma^2 \vc I)^{-1} \approx \vc I$).
Thus the predictive divergence at training points can go small even though the submodel predicts poorly.

This flexibility of the GP makes the formulation of the projection framework for GPs nontrivial, the problem being that it is easy to minimize the divergence in a finite number of observed training points, but difficult to ensure the generalization to new locations.
Based on the numerical experiments (Sec.~\ref{sec:experiments}), the proposed method appears to solve this ill behaviour to some extent, but it seems that there might still be room for improvement (see also the discussion in Sec.~\ref{sec:discussion}).

\subsubsection{Computational details}

For hyperparameter optimization, the objective function~\eqref{eq:gpproj_problem} becomes (leaving out the multiplier $1/2$ from Gaussian KL)
\begin{align*}
	E = \; & \tr{\vs \Sigma_\perp^{-1}\vs \Sigma} + 
	(\vs \mu  - \vs \mu_\perp)^\tp \vs \Sigma_\perp^{-1} (\vs \mu - \vs\mu_\perp)\\
	& + \log|\vs \Sigma_\perp| - \log|\vs \Sigma| - n.
\end{align*}
The derivative w.r.t. submodel parameter $\theta_\perp$ becomes
\begin{align*}
	&\frac{\partial E}{\partial \theta_\perp} =  
	\tr{ \frac{\partial \vc \Sigma_\perp^{-1}} {\partial \theta_\perp}
		 (\vs \Sigma - \vs \Sigma_\perp ) } + 
	\vs \mu^\tp \frac{\partial \vc \Sigma_\perp^{-1}} {\partial \theta_\perp} \vs \mu 
	 - \vc v^\tp \frac{\partial \vc \Sigma_\perp^{-1}} {\partial \theta_\perp} \vc v, \\
	&\frac{\partial \vc \Sigma_\perp^{-1}} {\partial \theta_\perp} = 
	-\vc K_\perp^{-1} \frac{\partial \vc K_\perp}{\partial \theta_\perp} \vc K_\perp^{-1}, \quad 
	\vc v = \sigma^{-2} \vc \Sigma_\perp \vc y.
\end{align*}
The computational complexity of the optimization is $O(n^3)$ and is dominated by the unavoidable matrix inversions.
Computations can be reduced, for example, to $O(m^2n)$ with sparse approximations using $m$ inducing points.
Thus the scaling is the same as for the usual hyperparameter optimization (maximum marginal likelihood), but in practice our method is slightly slower due to some matrix products not present in the marginal likelihood computation \cite{rasmussen2006book}.

\subsubsection{Non-Gaussian observation model}
\label{sec:gpproj_nongaussian}

In this paper we focus on the Gaussian observation model, but the approach presented in Section~\ref{sec:gpproj_gaussian} is easily extended to non-Gaussian observation models as it deals only with the latent function values, which can be approximated by Gaussian distribution.
For instance, in the case of binary classification, we can use expectation propagation (EP) \cite{minka2001} to approximate the posterior inference. 
In this case the observed outputs $\vc y$ and noise covariance $\sigma^2 \vc I$ in the posterior mean and covariance (in Eq.~\eqref{eq:gpproj_problem}) are replaced by the site parameters $\vs {\ti \mu}$ and $\vs {\ti \Sigma}$ \cite[p.~52--56]{rasmussen2006book}.

\subsubsection{Integration over the hyperparameters}
\label{sec:gpproj_hyperparams}

Typically we would like to account for uncertainty regarding the hyperparameters in the full model, for instance by sampling instead of optimizing them.
In principle we could use the original approach (Sec.~\ref{sec:projection_glm}) and do the projection separately for each posterior sample in the full model, but this is computationally expensive, because it multiplies the computation time by the number of samples.
Thus instead, we fix the noise variance $\sigma^2$ of the submodel to the posterior mean of the full model, compute the integrated mean and covariance $\vs \mu, \vs \Sigma$ of the full model by employing all the parameter values, and then find a single parameter vector for the submodel that best matches the resulting full model latent distribution.
This does not increase the computational time (since $\vs \mu$ and $\vs \Sigma$ must be computed only once), but accounts for the uncertainty regarding the hyperparameters to some extent.

\section{Experiments}
\label{sec:experiments}

\begin{figure}
	\centering 
	\setlength{\figureheight}{0.15\textwidth}
	\setlength{\figurewidth}{0.45\textwidth}
	\pgfplotsset{
	compat=newest,
	major tick length={0.05cm},
	legend style={font=\footnotesize}, 
	x tick label style={font=\footnotesize},
	y tick label style={font=\footnotesize},
	} 
	\minput[pdf]{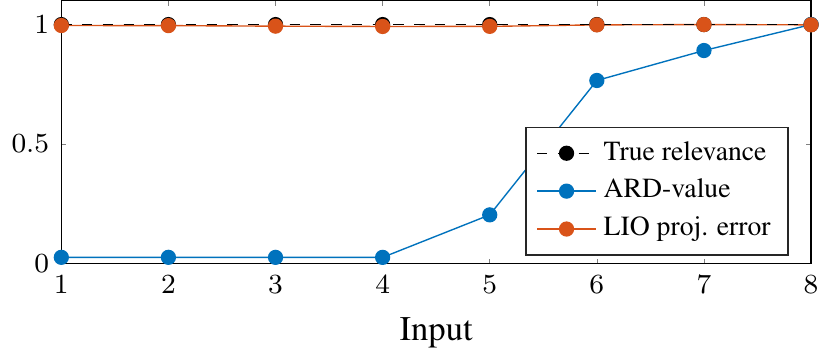}
	\caption{ARD values (inverses of the length-scales) and leave-input-out (LIO) projection errors for the toy example from Sec.~\ref{sec:ard}, averaged over 200 data realizations. The values are scaled so that the highest value is one. The projection error does not favor non-linear inputs like ARD.}
	\label{fig:ard2}
\end{figure}

\begin{table}%
\centering
\abovetopsep=2pt
\caption{Summary of the real world datasets: number of input variables $p$, dataset size $n_\tx{tot}$ and points used for training $n$. The rest of the points were used for testing.}
\label{tab:datasets}
\begin{tabular}{ lccc }
\toprule
Dataset & $p$ & $n_\tx{tot}$ & $n$ \\ 
\midrule
Boston housing &  13 & 506 & 300  \\ 
Automobile & 38 & 193 & 150  \\
Crime & 102 & 1992 & 400 \\
\bottomrule
\end{tabular}
\end{table}

\subsection{Toy example revisited}
\label{sec:toy_example_revisited}

Let us briefly revisit the toy example from Sec.~\ref{sec:ard} and demonstrate how the projection performs in this case.
Employing the fitted full model with hyperparameters fixed to the maximum marginal likelihood, we computed the leave-input-out projection errors for each input, that is, divergences for all models missing one input.
Figure~\ref{fig:ard2} shows the results averaged over the 200 data realizations.
The results show that, on average, the removal of each input introduces the same projection error, demonstrating that the projection does not favor the nonlinear inputs like the ARD does and gives a more realistic assessment of the input relevances.

\subsection{Real world data}

\begin{figure*}
	\centering 
	\setlength{\figureheight}{0.13\textwidth}
	\setlength{\figurewidth}{0.9\textwidth}
	\pgfplotsset{
	compat=newest,
	major tick length={0.05cm},
	legend style={font=\footnotesize}, 
	x tick label style={font=\footnotesize},
	y tick label style={font=\footnotesize},
	} 
	\minput[pdf]{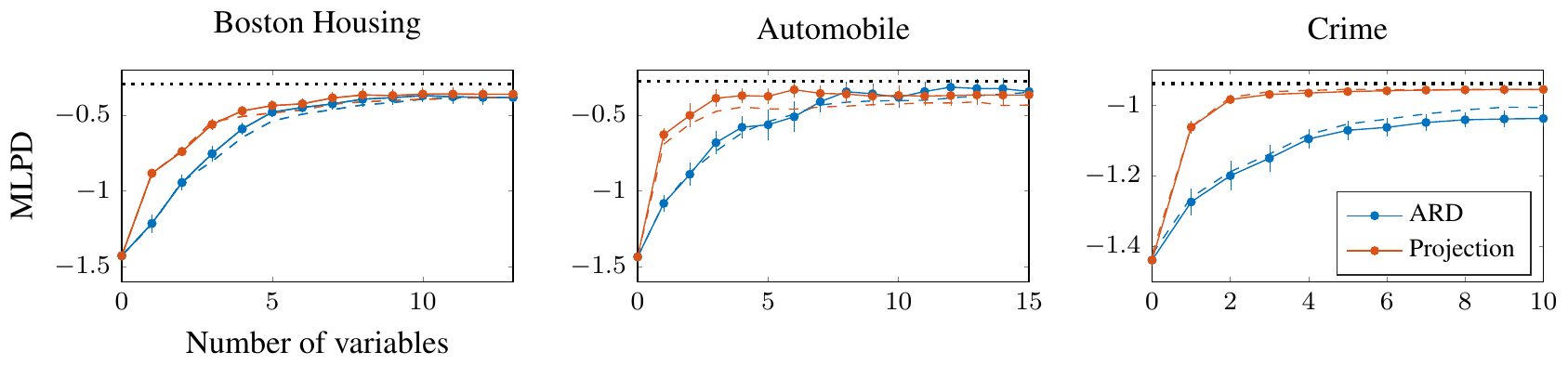}
	\caption{Mean log predictive density (MLPD) on the test set with 95\%-intervals for the submodels as a function of number of inputs added, when the variables are sorted by ARD (blue) or by stepwise forward searching minimizing the projection error from the full model (red). The solid lines indicate that the submodel parameters are learned by projection, and dashed line that they are optimized to the maximum marginal likelihood. The horizontal dotted line is the accuracy of the full model with sampled hyperparameters.}
	\label{fig:real_data}
\end{figure*}

We also tested the performance of our method on three public datasets from the UCI machine learning repository\footnote{\url{http://archive.ics.uci.edu/ml/}}\footnote{The datasets required some preprocessing, and for Automobile data we used the price as the target variable to obtain a regression problem.}, summarized in Table~\ref{tab:datasets}.
We ran the projective variable selection for 50 random splits into training and test sets, using the training set for selection and then studied the predictive performance on the independent test data as a function of number of variables selected.
As a search heuristic, we used the stepwise forward searching, i.e., at each step adding the input that decreases the projection error~\eqref{eq:pred_divergence} the most.
We used constant and squared exponential~\eqref{eq:se_kernel} covariance functions, and the full model was constructed by drawing 100 samples from the hyperparameter posterior (with uniform priors for the logarithms of the hyperparameters) using Hamiltonian Monte Carlo (HMC), to account for hyperparameter uncertainty.
The results were compared to the ARD, that is, optimizing the hyperparameters to the maximum marginal likelihood and sorting the inputs based on their length-scales.

Figure~\ref{fig:real_data} shows the results.
For all three datasets the projection finds models with predictive accuracy close to the full model with fewer inputs compared to ARD, the difference being clearer for the Automobile and Crime datasets.
For Automobile dataset, learning the submodel parameters by projection (solid lines) appears to improve the predictive accuracy slightly compared to optimizing them to the maximum marginal likelihood (dashed line), but overall the difference is rather small.
Thus, the difference is mainly related to improved ordering of the variables.
Nevertheless, even with the projection the predictive accuracy does not converge to the level of the full model, suggesting that projecting only a point estimate of the parameters does not fully capture the uncertainties in the full posterior of the parameters (see Sec.~\ref{sec:gpproj_hyperparams}).
Moreover, the price paid for better ordering of the variables is a substantial increase in computations; while ARD does not require any additional searching once the full model is fitted, the forward search for projection takes about 40 minutes for Boston housing and 4 and 15 hours for Automobile and Crime datasets (see Sec.~\ref{sec:discussion} for discussion).

\section{Discussion}
\label{sec:discussion}

We have proposed a new method for simplification of GPs based on the projection predictive approach, originally proposed \cite{goutis1998,dupuis2003} and shown to be useful for generalized linear models \cite{piironen2016}. In this paper we extended the approach for non-parametric GP models.
The results on synthetic and real world data show that the our new method provides a more accurate assessment of the input relevances than the often used ARD values (inverse length-scales) which tend to favor inputs with nonlinear response.
More precisely, our method finds submodels with predictive accuracy close to the full model with fewer inputs than sorting the variables based on their length-scales.

The method proposed here requires use of some search heuristic, such as the forward search used in our experiments, to find promising submodels with fewer variables.
This is computationally demanding even for relatively small problems, as one has to fit several GPs along the search. 
The experiments also suggested that projecting only a single parameter vector does not fully capture the uncertainties presented by the full model posterior, even when all the variables have been added (Boston data).
Projecting several parameter samples separately would provide improved predictive accuracy, but would make the search even slower. %
The method could be made faster by sparse approximations, greedy search methods and additional computational heuristics.
Despite these issues, we believe this work is a step towards more robust and accurate model simplification for GPs. 

\bibliographystyle{IEEE}
\bibliography{references}

\end{document}